\documentstyle[11pt]{article}
\begin{document}
\def\be{\begin{equation}}
\def\en{\end{equation}}
\def\bear{\begin{eqnarray}}
\def\enar{\end{eqnarray}}
\arraycolsep 2pt
\newcommand{\square}
{\kern1pt\vbox{\hrule height 0.9pt\hbox{\vrule width 0.9pt\hskip 3pt
\vbox{\vskip 6pt}\hskip 3pt\vrule width 0.6pt}\hrule height 0.6pt}\kern1pt}
\title{\bf Particle Creation in the Bell-Szekeres Spacetime}
\author{\it Alexander Feinstein and\\
        \it Miguel Angel P\'erez Sebasti\'an\\
        \it Dpto. F\'\i sica Te\'orica, Universidad del Pa\'\i s Vasco,\\
        \it Bilbao, Spain.}
\date{}
\maketitle
\vskip 1.5cm
\begin{abstract}
The quantization of a real massless scalar field in a spacetime  produced in a
collision of two electromagnetic plane waves with  constant wave fronts is
considered. The background geometry in  the interaction region, the
Bell-Szekeres solution, is locally  isometric to the conformally flat
Bertotti-Robinson universe filled  with a uniform electric field. It is shown
that before the waves  interact the Bogoliubov coefficients relating different
observers  are trivial and no vacuum polarization takes place. In the non-
singular interaction region neutral scalar particles are produced  with number
of created particles and spectrum typical of gravitational wave collision.
\end{abstract}
\newpage

\section {Introduction}

It is known that quantum particles are not produced in the  vicinity of plane
electromagnetic waves due to the high degree of symmetry
\cite{schwinger,deser}.  Neither, one expects particle creation when such waves
scatter in  a flat spacetime, without gravity taken into account, because of
the  linearity of the process. If the electromagnetic waves are
self-gravitating,  however,  they interact non linearly producing a region with
a non zero Coulomb  component of the electric field.  One then expects
particles with charge to be  created.  Thus, the effect of quantum particle
creation when two self- gravitating electromagnetic waves scatter can be
considered as a purely  curvature effect caused by the non-linear interaction
of the fields governed by the  coupled Einstein-Maxwell equations.

The simplest example of the collision of the plane electromagnetic  waves was
given by Bell and Szekeres \cite{BS} and studied later by Matzner  and Tipler
\cite{MT} and by Clarke and Hayward \cite{CH}.   The example considered by Bell
and Szekeres involves a collision  of two step plane waves with constant
wave-fronts which, unlike most of the  waves after the collision, do not focus
to a curvature singularity, but rather, to  a Killing-Cauchy horizon.   It was
shown by Clarke and Hayward further on, that the solution  in the interaction
region is extendible across the focusing surface  similarly to the previously
studied cases of non-singular collisions of pure  gravitational plane waves
\cite{FI,yurtsever}.

Matter field quantization on the colliding wave background was  first studied
by Yurtsever \cite{yurtsever2} with the background produced by  the collision
of
two plane impulsive gravitational waves, the Khan-Penrose  solution \cite{KP}.
Because of the peculiar property of the Khan-Penrose solution  with flat
regions in the single-wave propagating parts of the spacetime,  Yurtsever has
managed to construct, in a relatively simple way, an unambiguous  ``out"-vacuum
related to these flat regions behind wave fronts.  Generally, however, the
spacetime regions behind the wave fronts  have nonzero curvature and it is not
easy to construct the ``out"-modes with the  procedure outlined by Yurtsever.

Dorca and Verdaguer \cite{DV, DV2} have noticed recently that the  presence of
the Killing-Cauchy horizon, instead of the strong curvature  singularity, makes
the task of solving the quantum field theory on the background of  colliding
waves technically more plausible.  The presence of the Killing-Cauchy horizon
and the symmetries  associated with it can be used to define the unique
preferred vacuum state, as   pointed out by Kay and Wald \cite{KW} in a general
context.

In this paper we consider the quantization of massless {\it  neutral} scalar
field on the background geometry produced by the collision of  step
electromagnetic plane waves with constant wave fronts - the Bell- Szekeres
solution.  The interaction region of the Bell-Szekeres solution is  isometric
to the Bertotti-Robinson universe \cite{BR,gribook} - the static  conformally
flat solution of the Einstein-Maxwell equations with uniform  electric field.

Quantum test electrodynamics on the Bertotti-Robinson  background, within the
context of the Euclidean quantum field theory, was previously  considered by
Lapedes\cite{lapedes}.  The Bertotti-Robinson universe has three non-commuting
timelike  Killing vectors and one may associate three different observers
related to these  vector fields experiencing different accelerations
\cite{lapedes}. The  observers  ``see" different spectra of created particles
according as to  whether the acceleration experienced by the observers exceeds
a certain  critical value $a_{cr}$. If the acceleration exceeds $a_{cr}$ then
the spectra of  the created particles has a form typical to a Hawking thermal
spectrum, whereas if the acceleration is equal or smaller than the $a_{cr}$ the
spectrum  is nonthermal approaching the Schwinger spectrum \cite{schwinger}  in
the limit of high electric field strength.

In this work we are interested in the creation of neutral scalar pairs, rather
than charged particles, because the production of the  latter ones would not
come in as a surprise in a region of spacetime (the  interaction region of the
Bell-Szekeres solution) which can be thought of as  filled with  uniform
electric field. The effect of neutral particle creation in the Bell-Szekeres
solution, however, can be considered as not only due  to the nonlinear
interaction of the waves as stated at the beginning but  also as a result of
the dynamical evolution of the spacetime.

While maybe somewhat simplified, the Bell-Szekeres example of the scattering of
two electromagnetic  waves represents an interesting theoretical laboratory to
study the  quantum field theory.  This is mainly due to the simplicity of the
metric in each  of the different regions defined in the problem of plane wave
collision.  Nevertheless, the global compositeness of the spacetime preserves
all the features of more complicated problems involving plane wave  scattering.

In the following Section 2 we briefly discuss some relevant  geometrical
properties of the Bell-Szekeres solution defining different  coordinate patches
to be used in this paper.  In Section 3 we solve exactly the Klein-Gordon
equation in all four regions associated with the wave collision.  We also
discuss the nonexistence of vacuum polarization in the  case of a single plane
electromagnetic wave.  We then follow Dorca's and Verdaguer's idea \cite{DV}
propagating the ``in"-vacuum state from the initially Minkowskian background
into the interaction region via the regions of single plane waves and then
define the ``out"-vacuum relating it to the observer at rest at the Killing-
Cauchy horizon.   In Section 4 we evaluate the Bogoliubov coefficients relating
the  ``in" and the ``out" modes and calculate the number of particle created as
seen by the observer at the horizon.

\section {Geometric Properties}

The Bell-Szekeres solution represents the collision of two  electromagnetic
plane waves with different amplitudes, and constant polarization
\cite{BS,gribook}.

The metric tensor for this spacetime is:
\be
ds^{2}=2du dv-\cos^{2}[au\theta(u)-bv\theta(v)] dx^{2}-
\cos^{2}[au\theta(u)+bv\theta(v)] dy^{2}
\en
where $au<\pi/2,\; bv<\pi/2,\; -\infty <x,y<\infty; \;$ here the  positive
constants $a$ and $b$ are related to the strengths of the  electromagnetic
plane waves.

The metric coefficients have square-integrable weak derivatives,  so that the
curvature tensor can be split into a regular and  distributional parts:
\be
R^{\alpha}_{\beta\gamma\delta} = \hat R^{\alpha}_{\beta\gamma
\delta}+\overline R^{\alpha}_{\beta\gamma\delta}
\en
Here the distributional part $\overline  R^{\alpha}_{\beta\gamma\delta}$  is a
linear combination of the following distributions: $\delta (u)\theta (v)\sin
(bv)$ and $\delta (v)\theta (u)\sin (au)$.  The regular part of the Ricci
tensor is zero, and the curvature scalar is zero globally:
\be
R = 0
\en
The components of the Weyl tensor as well as the electromagnetic tensor are
given in reference  \cite{CH}.
The spacetime represents two electromagnetic and gravitational impulsive
colliding plane  waves along \( u=0,\, v=0\) hypersurfaces. Note, that there
are no gravitational waves before the collision and these are induced  only
after the scattering has taken place.

The spacetime is split into four different regions:
\be
\begin{array}{rcl}
ds^{2}_{IV}&=&2du dv - dx^{2} - dy^{2},\quad\qquad\qquad u
, v < 0\\*[8pt]
ds^{2}_{III}&=&2du dv - \cos^{2}(bv)(dx^{2} + dy^{2}), \quad u
< 0 , 0
< v <\pi/2b\\*[8pt]
ds^{2}_{II}&=&2du dv - \cos^{2}(au)(dx^{2} + dy^{2}), \quad v <
0 , 0 <
 u <\pi/2a\\*[8pt]
ds^{2}_{I}&=&2du dv - \cos^{2}(au-bv)dx^{2} -
\cos^{2}(au+bv)dy^{2}, \; au + bv <\pi/2.
\end{array}
\en

\noindent The metric is regular in each region with apparent singularities in:
($u={\pi\over 2a},\; v<0$),
($v={\pi\over 2b},\; u<0$), and
($u>0,\; v>0,\; au+bv={\pi\over 2}$)
which can be removed by a change of coordinates. The first two  are fold
singularities and the last one is the Killing-Cauchy horizon where  the plane
waves focalize \cite{BS,CH}.

Region IV is Minkowski spacetime. One can show that regions II  and III are
conformally flat by performing the following coordinate  transformation (in
the region II, for example):
\be
\bar u = \tan au \label{conftranII}
\en
then the metric in region II takes the form:
\be
ds^2={1\over 1+\bar u^2}\;(\frac{2}{a}\; d\bar u\, dv - dx^2 -
dy^2)
\label{metIIconf}
\en

Region I is also conformally flat.  It can be brought to the  explicitly
conformally flat form by the following coordinate transformation
\cite{griffiths}:
\be
\begin{array}{rcl}
t + r & = & \coth [{1\over 2}\;{\rm sech}^{-1}(\cos(au+bv))-
{y\over 2q}]\\*[8pt]
t - r & = & -\tanh [{1\over 2}\;{\rm sech}^{-
1}(\cos(au+bv))+{y\over 2q}]\\*[8pt]
\theta & = & \pi/2 -(bv-au) \label{conftrans}\\*[8pt]
\phi & = & x/q,
\end{array}
\en

\noindent where $q={1\over\sqrt{2ab}}$. Then, the metric tensor is simply:
\be
ds^2={q^2\over r^2}(dt^2-dr^2-r^2d\theta^2-r^2\sin^2\theta
d\phi^2). \label{b-r}
\en
One can immediately recognize the line element (\ref{b-r}) as the
Bertotti-Robinson  spacetime which has a geometry similar to that  one of the
throat of the Reissner-Nordstrom solution for the special case  $Q=M$
\cite{MTW}.

In the Bertotti-Robinson solution, the coordinate $\phi$ is cyclic:
$0\le\phi<2\pi$, while in the Bell-Szekeres solution, however, the
corresponding $x$ coordinate is defined over the whole range:
$-\infty<x<\infty$.

The Bertotti-Robinson coordinates are not well adapted to  describe the
Killing-Cauchy horizon. It is convenient, therefore, to introduce a  new set of
coordinates in order to describe the interaction region.  One then defines a
set of Kruskal-Szekeres-like coordinates in the  following way \cite{DV}:

First we define a new dimensionless time and space coordinates  $(\xi,\eta)$:
\be
\begin{array}{rcl}
\xi & = & au+bv \qquad 0\le\xi <\pi/2\\*[8pt]
\eta & = & bv-au \qquad -\pi/2\le\eta <\pi/2\label{xieta},
\end{array}
\en

\noindent introduce a new dimensionless time-like coordinate $\xi^\ast$:
\be
\xi^\ast =
{1\over\sqrt{2ab}}\,\log\left({1+\sin\xi\over\cos\xi}\right)
\label{xiasterisco},
\en
and a new set of null coordinates:
\be
\begin{array}{rcl}
\tilde U&=&\xi^{\ast} - y\\*[8pt]
\tilde V&=&\xi^{\ast} + y \label{defUV},
\end{array}
\en

\noindent Finally, we define the Kruskal-Szekeres-like null coordinates:
\be
\begin{array}{rcl}
U'=-q\, e^{-\tilde U/q}\\*[8pt]
V'=-q\, e^{-\tilde V/q} \label{defU'V'}
\end{array}
\en

\noindent so that the metric in the interaction region becomes:
\be
ds^2=(1+\sin\xi)^2\;dU'\,dV' - {1\over 2ab}d\eta^2
-\cos^2\eta\;dx^2,
\en
with
\be
U'\,V' = q^2\, {1-\sin\xi\over 1+\sin\xi},
\en
and
\be
{U'\over V'} = e^{2y/q}.
\en
The curves $\xi=const.$ are hyperbolae and $y=const.$ are  straight lines in
the  $(U',V')$ plane. When $\xi\to\pi/2$ we obtain hyperbolae
$U'\,V'=\epsilon$, $\epsilon >0$. And the hypersurface which is  the
Killing-Cauchy horizon $\xi=\pi/2$ is then $\{ (U'=0, V'\le 0) \cup   (V'=0,
U'\le 0)\}$.

\section {Quantization of the scalar field}

The quantization of a scalar
field and the production of particles on a given curved  background is done
in a standard manner (see for example \cite{BD}). We will proceed as follows:
first construct a complete orthonormal  set of modes for a  massless real
scalar field related to the  Minkowskian region IV of the  spacetime before the
collision.  In this region, one is able to build a Fock space  related to an
inertial observer.  These modes will be consider as the ``in"-modes  and,
since the region is flat, all the inertial observers in this  region  would
agree on the definition of particles \cite{BD}.

Next, we propagate these modes throughout all the spacetime up  to horizon  by
solving the Klein-Gordon equation in each region of the  spacetime and matching
the modes across different hypersurfaces separating the  regions.  At each
state we will be able to solve exactly the Klein-Gordon  equation in every
region of spacetime.

In passing, we study the case of a single electromagnetic plane  wave. Using
two different observers, one related to the ``in"-vacuum state  propagated from
the region IV into the region II and the other which is  related to the modes
constructed using the conformal symmetry of the single  wave region we
explicitly show that the Bogoliubov coeficients are trivial and  there is no
particle creation in the vicinity of the plane wave. The triviality  of these
Bogoliubov coeficients prevents to construct a different set of  modes and to
proceed the quantization in the interaction region in the way done  by
Yurtsever \cite{yurtsever2}. Neither it is simple to define a  different set of
modes using the harmonic coordinates due to the nonflatness of  the single wave
region. One therefore is bounded to use the procedure of  Dorca and Verdaguer
\cite{DV} which is most suitable to our case due to the  presence of two null
Killing vector fields at the horizon.

Formally, we could have tried to use different symmetries  associated with the
geometry of the interaction region: the conformal flatness or the  existence of
three non-commuting timelike Killing vectors \cite{lapedes}.  This, however,
leads to serious difficulties with the definition of particles due to
compositness of the Bell-Szekeres spacetime which limits the  range of the time
coordinate related to these symmetries.

\subsection {Region IV}

The metric in the region IV is the Minkowski spacetime:
\be
ds^2={2\over ab}dudv-dx^2-dy^2 \quad u<0,\,v<0,
\en
where we have rescaled the coordinates:
\bear
u' & = & a u \\
v' & = & b v, \label{adimnull}
\enar
and then abolished the primes.

The general solution of the Klein-Gordon equation for a massless scalar
field can be expanded into the following orthonormal set of modes:
\be
u^{in IV}_k(u,v,x,y)={1\over\sqrt{2k_-(2\pi)^3}}\exp{\{-i{k_+
\over a}u-i{k_-\over b}v+ik_xx+ik_yy\}}\label{iniv}
\en
where  $\; k_-k_+ = 1/2\, (k_x^2+k_y^2)$.

In spite of the fact that the single plane wave spacetimes do not  contain  a
global Cauchy surface \cite{penrose} in order to construct the  scalar product,
it was shown by Gibbons \cite{gibbons} that one can use  the null surfaces
$u=const$ instead. One can then see that the modes given by the equation
(\ref{iniv}) are well   normalized on the ``roof" hypersurface $\Sigma
:\big\{(u=0,v<0)\cup(u<0,v=0)\big\}$:
\be
\big(u^{in IV}_k,u^{in IV}_{k'}\big)=\delta (k_x-k'_x)\delta(k_y-
k'_y)
\delta(k_--k'_-).
\en
These modes impose the following boundary conditions on the  ``in"-modes in the
single wave region II (all expressions in III can be obtained by
interchanging $u \leftrightarrow v$, $a \leftrightarrow b$ and $k_-
\leftrightarrow k_+$) due to the continuity:
\be
u^{in II}_k\big\vert_{u=0}=
u^{in IV}_k\big\vert_{u=0}=
     {1\over\sqrt{2k_-(2\pi)^3}}\;\exp{\{-i{k_-\over
b}v+ik_xx+ik_yy\}}
\label{boun1}
\en
\subsection {Regions II and III.-}

The metric in the region II takes the following form:
\be
ds^2={2\over ab}dudv-\cos^2 (u)(dx^2+dy^2) \quad
v<0,\,0<u<{\pi\over 2}
\en
Representing the passage of a single electromagnetic  plane wave.

The corresponding Klein-Gordon equation for the  region II becomes:
\be
\phi_{uv}\,-\,\tan u \,\phi_v\,-\,{\phi_{xx} + \phi_{yy}\over
2ab\cos^2u} =0
\label{kgII}
\en
For the region III the corresponding Klein-Gordon equation is basically the
same equation as (\ref{kgII}) changing $u$ for $v$.   The  ``in"-mode solutions
for the region II satisfying the boundary  conditions  (\ref{boun1}) are:
\be
u_k^{in II}(u,v,x,y)={1\over\sqrt{2k_-(2\pi)^3}}\,f(u)
\,\exp{[-i{k_-\over b}v+ik_xx+ik_yy]}
\label{modesinII}
\en
where $f(u)$ is:
\be
f(u)={1\over\cos (u)}\exp{[-i{k_+\over a}\tan (u)]}.
\en

We now choose the null hypersurface $\Sigma:\left\{ (v=0,\, 0\le  u<\pi/2)\cup
(u=0,\,\right.$ $\left. 0\le v<\pi/2)\right\}$, which is the  characteristic
surface of Klein-Gordon equation in these regions, to  orthonormalize the
``in"-modes:
\bear
(u^{in}_k,u^{in}_{k'})&=&-i\int dxdy\int_0^{\pi/2}\cos^2(u)
\big(u^{in II}_k\stackrel{\leftrightarrow}{\partial}_u u^{in II
\ast}_{k'}\big)\bigg\vert_{v=0} du \\
& &-i\int dxdy\int_0^{\pi/2}\cos^2(v)
\big(u^{in III}_k\stackrel{\leftrightarrow}{\partial}_v u^{in III
\ast}_{k'}\big)\bigg\vert_{u=0} dv,\nonumber
\enar
which gives:
\be
\big(u^{in}_k,u^{in}_{k'}\big)=\delta(k_x-k'_x) \delta(k_y-k'_y)
\delta(k_--k'_-).
\en
The ``in"-modes then are orthonormal considering these two  propagation
regions, and induce, in turn, the following boundary conditions for  the
``in"-modes in the interaction region across the hypersurfaces:
\be
u^{in I}_k\vert_{v=0}=u^{in II}_k\vert_{v=0}={1\over\sqrt{2k_-
(2\pi)^3}}{1\over \cos (u)}\exp{[-i{k_+\over a}\tan (u)+ik_xx+ik_yy]}
\label{bound}
\en
And the corresponding for the $u=0$ hypersurface related to the ``in"-modes in
the region III.

Although the normal ``in"-modes in the region II (III) diverge
at  the points  of the fold singularities ($\{ u=\pi/2, v=0\}$, $\{ v=\pi/2,
u=0\}$),  this  divergence does not influence the scalar product in this
region  because, as argued by Dorca and Verdaguer \cite{DV} only a set of null
measure of these modes arrive at these points.

\subsection {Conformally flat modes in the single wave region}

We have seen that the regions II and III corresponding to the  propagation of
single electromagnetic waves, are conformally flat. Therefore, an  observer
adapted to the symmetries of the region i.e., an observer who sees  the region
as a conformally flat one would be a physically meaningful  observer (see for
example \cite{BD}). Following the coordinate change which  transforms the
metric into an explicitly conformally flat form for the region II  (for
example):
\be
\bar u = \tan u, \label{ubarra}
\en
the metric becomes:
\be
ds^2 = {1\over 1+\bar u^2} \left( {2\over ab}\, d\bar u\, dv -  dx^2
- dy^2\right) = \Omega^2 (u)  \left( {2\over ab}\, d\bar u\,
dv -
dx^2 - dy^2\right)
\en
The solutions for a massless scalar field of the corresponding  Klein-Gordon
equation in our case are simply the plane waves with the amplitude
multiplied by the inverse of the conformal factor:  $\Omega (u)$
\be
u_k^{conf} (\bar u,v,x,y) = {\sqrt{1+\bar u^2}\over\sqrt{(2\pi)^3
2k_-}}\,
\exp\left[-i{k_-\over b}v -i{k_+\over a}\bar u
+ik_xx+ik_yy\right],
\label{confII}
\en
where the coeficients are related by $k_-k_+ = (k_x^2 +k_y^2)/2$.  The
conformal modes for the region III are obtained by changing  ($a,\bar u$) by
($b,\bar v$).

We now look at the Bogoliubov transformation between these  conformal modes
(\ref{confII}) and the ``in"-modes (\ref{modesinII}) in the  single wave
regions :
\be
u^{in}_k = \sum_{k'} \left(\alpha_{kk'}\,u^{conf}_{k'} +
                            \beta_{kk'}\,u^{conf\ast}_{k'} \right).
\en
After some simple algebra, introducing the inverse of the  transformation
equation (\ref{ubarra}) in the definition of the conformal modes
(\ref{confII}), it can be easily shown that the Bogoliubov  coeficients are
trivially:
\be
\alpha_{kk'} = \delta^3(k-k'),\quad\beta_{kk'}\equiv 0
\en
So that, the two sets of modes are in fact the same set.

This implies that there is no particle creation in the single  electromagnetic
wave region, which is in full agreement with earlier studies  (\cite{deser},
\cite{gibbons}, \cite{GV}, \cite{LP}). Also due to a particular case of the
electromagnetic wave the ``conformal" observer and  the ``in" observer that
arrives from the Minkowskian region have the same  definition of particles.

\subsection {Interaction region.-}

The metric tensor in the interaction region is given by:
\bear
ds^2={2\over ab}dudv-\cos^2(u-v)dx^2-\cos^2(u+v)dy^2 \\
u>0,\;v>0,\; u+v\le\pi/2 \nonumber
\enar
In these coordinates the Klein-Gordon equation for the scalar field  is non
separable. We thus change to the following dimensionless  coordinates:
\bear
\xi & = & u+v \nonumber \\
\eta & = & v-u, \nonumber
\enar
and the line element becomes:
\be
ds^2={1\over 2ab}d\xi^2-{1\over 2ab}d\eta^2-\cos^2\eta\,dx^2-
\cos^2\xi\,dy^2
\label{intmetric}\\
\en
The Klein-Gordon equation in this region then reads:
\be
2ab\Phi_{,\xi\xi}-2ab\Phi_{,\eta\eta}-2ab\tan\xi\,\Phi_{,\xi}+2ab
\tan\eta\,\Phi_{,\eta}-{\Phi_{,xx}\over\cos^2\eta}-{\Phi_{,yy}
\over\cos^2\xi} = 0.
\en
It is convenient to separate the solution in the following form:
\be
\Phi (\xi,\eta,x,y) =e^{ik_yy}\varphi (\eta,x)\psi (\xi),
\en
obtaining two decoupled differential equations:
\bear
\varphi_{\eta\eta}-\tan\eta\,\varphi_{\eta}+{\alpha\over
2ab}\,\varphi
+{\varphi_{xx}\over\cos^2\eta} & = & 0\label{etax}\\
\ddot\psi-\tan\xi\,\dot\psi+\bigg\{{\hat k_y^2\over\cos^2\xi}+
{\alpha\over 2ab}\bigg\}\psi & = & 0 \label{xi}.
\enar
Here $\alpha$ is the separation constant and $\hat
k_i=k_i/\sqrt{2ab}$.

The solutions of the first differential equation (\ref{etax}) are  given by the
product of exponentials and associated Legendre functions:
\be
\varphi (x,\eta)\propto e^{ik_xx}\,P^{\hat k_x}_\mu (\cos\eta),
\en
where $\mu$ is defined by ${\alpha\over 2ab}=\mu(\mu + 1)$.

We now identify the coordinate $x$ with the angular coordinate  with the range:
$0<x\le 2\pi L$ (see also \cite{DV} and \cite{hayward}),  and imposing the
regularity conditions on the axis one can show that $L=q$  independently on
which possible analytic extension one wishes to perform across  the
Killing-Cauchy horizon. The solution thus can be written in  spherical harmonic
functions:
\be
\varphi_{l,m}(\eta)\propto Y^{m}_l({\pi\over 2}-\eta,{x\over q})
\en
where $m\equiv\hat k_x$ and $l\equiv\mu$ are integers
related in the usual form ($l=0,1,..\infty ;\; m=-l,..,l$).

The second equation can be put into the hypergeometric form by  the following
transformations:
\bear
z&=&\sin\xi\\
\psi(z)&=&(1-z^2)^{i\hat k_y}\,\phi(z)\\
u&=&{1\over 2}(1-z),
\enar
we then have:
\be
u(1-u)\ddot\phi+(1+i\hat k_y)(1-2u)\dot\phi+({k_y^2+l(l+1)\over
2ab}-i\hat k_y)
\,\phi=0\label{hyp}.
\en
Using the properties of the hypergeometric functions, the general  solution of
the equation (\ref{hyp}) is a linear combination of the following  solutions:
\be
\psi_{l,\hat k_y,1}(\xi) =
\bigg({1+\sin\xi\over 1-\sin\xi}\bigg)^{-i\vert \hat k_y\vert
/2}\;
 _2F_1\bigg[1+l, -l;1+i\hat k_y;{1\over 2}(1-
\sin\xi)\bigg],\nonumber
\en
and
\be
\psi_{l,\hat k_y,2}(\xi) =
\bigg( {1+\sin\xi\over 1-\sin\xi}\bigg)^{i\vert \hat k_y\vert /2}\;
_2F_1\bigg[1+l, -l;1-i\hat k_y;{1\over 2}(1-\sin\xi)\bigg]
\label{intsol}
\en
The general solution in the interaction region finally is:
\be
\Phi(\xi,\eta,x,y)=e^{ik_yy}\sum_{l}Y^{m}_l({\pi\over 2}-\eta,
{x\over q})
\big( C^{(1)}_l\psi^{(1)}_{l,k_x}(\xi)+C^{(2)}_l\psi^{(2)}_{l,k_x}
(\xi)\big)\label{general}
\en
The coeficients $C^{(1)}$ and $C^{(2)}$ depend on the separation  constant and
are subject to the boundary conditions (\ref{bound}).

\subsubsection {The ``in"-modes near the horizon}

It is usually difficult to define the ``out"-modes in the interaction  region
of the general colliding plane wave spacetime. Even in our case, in  spite of
the presence of certain symmetries, the definition of the ``out"- modes is
rather complicated. Fortunately however, as pointed out by Dorca  and Verdaguer
\cite{DV}, the existence of a Killing-Cauchy horizon helps one to define an
unambiguous ``out"-vacuum. One thus is  interested in the  asymptotic behaviour
of the ``in"-modes at the horizon.

To study the behaviour of the ``in"-modes near the horizon we can  proceed in
two different manners. We can either look at  the asymptotic form  of the
equation (\ref{xi}) near the horizon and then solve it or look  directly
at the asymptotic behaviour of the solutions given by the equation
(\ref{general}) at the horizon.  Both give the same result. The  equation
(\ref{xi}) can be written as:
\be
\psi_{\xi^\ast\xi^\ast} +
(k^2_y+\alpha\cos^2\xi)\psi = 0\label{xiast}
\en
where $\xi^\ast$ is defined by (\ref{xiasterisco}).

Near the horizon at $\xi\to\pi/2$ this equation can be simplified  to an
oscillator equation:
\be
\psi_{\xi^\ast\xi^\ast} + k^2_y \psi = 0
\en
and the solutions are:
\bear
\psi_1(\xi^\ast)&\propto &\, e^{-i\vert
k_y\vert\xi^\ast}\nonumber\\
\psi_2(\xi^\ast)&\propto &\, e^{i\vert
k_y\vert\xi^\ast}\label{waves}
\enar
Here $\xi^\ast$ is a time coordinate, so that the first term  represents a
purely ingoing wave to  the horizon while the second term represents a purely
outgoing  wave.

If the solutions are expressed in the original $\xi$ coordinate,  these take
the form:
\be
       \psi_1(\xi\to \pi/2)\propto
       \bigg({1+\sin\xi\over 1-\sin\xi}\bigg)^{-i\vert \hat k_y\vert
       /2}
\en
\hskip 2cm
and
\be
       \psi_2(\xi\to \pi/2)\propto
       \bigg({1+\sin\xi\over 1-\sin\xi}\bigg)^{i\vert \hat k_y\vert
       /2}.
\label{asym}
\en
One can easily see that $\psi_1(\xi^\ast)$ and $\psi_2(\xi^\ast)$ of  the
equation (\ref{asym}) are the  asymptotic forms of  $\psi^{(1)}_{\mu,k_x}(\xi)$
and $\psi^{(2)}_{\mu,k_x}(\xi)$ of the equation (\ref{intsol}) near the horizon
respectively. We can also see that the potential in the equation (\ref{xiast})
vanishes at the horizon $\xi\to\pi/2$, which allows one \cite{DV} to consider
the ingoing  modes only  $(C^{(2)}=0)$.

In this region the ``in" modes are defined as:
\be
u_k^{inI}(\xi,\eta,x,y)={e^{ik_yy}\over\sqrt{(2\pi)2\vert
k_y\vert}}
\sum_{l}C^{(1)}_l\psi^{(1)}_{l,m}(\xi) Y^m_l({\pi\over 2}-
\eta,{x\over q}),
\label{inmodint}
\en
and near the horizon they behave as:
\be
u^{inI}_k(\xi\simeq\pi/2)={e^{ik_yy}\over\sqrt{(2\pi)2\vert
 k_y\vert}}\sum_lC^{(1)}_l e^{-i\vert k_y \vert\xi^{\ast}}
Y^m_l({\pi\over 2}-\eta,{x\over q}).
\en
In the ($\tilde U, \tilde V$) coordinates this expression takes the  form:
\be
u^{inI}_k(\xi\simeq\pi/2)={1\over\sqrt{(2\pi)2\vert k_y\vert}}
          \sum_lC^{(1)}_l Y^m_l({\pi\over 2}-\eta,{x\over q})
          \left\{
                 \begin{array}{ll}
                                 e^{-i\vert k_y\vert\tilde U}\; ,k_y\ge 0\\
                                 e^{-i\vert k_y\vert\tilde V}\; ,k_y\le 0
                 \end{array}
          \right.
\en
Finally, using the Kruskal-Szekeres-like null coordinates defined  by
(\ref{defU'V'}),  we obtain:
\be
u^{inI}_k(\xi\simeq{\pi\over 2})={1\over\sqrt{(2\pi)2\vert
k_y\vert}}
          \sum_lC^{(1)}_l Y^m_l({\pi\over 2}-\eta,{x\over q})
          \left\{
                 \begin{array}{ll}
                       (-U'/q)^{i\vert\hat k_y\vert}\, ,k_y\ge 0\\
                       (-V'/q)^{i\vert\hat k_y\vert}\, ,k_y\le 0.
                 \end{array}
          \right.\label{goodinmodes}
\en
The equation (\ref{goodinmodes}) defines the ``in"-modes in the
Kruskal-Szekeres-like   coordinates at the horizon.  These coordinates are
important because of their relation to the  null Killing  vector fields at the
horizon  which will be further used to define a  new set of normal modes.

The coeficients $C^{(1)}_l$ are subject to the following  orthonormalization
relation at the horizon:
\be
(u^{in}_k,u^{in}_{k'}) = q^2\,\delta(k_y-k'_y)\delta_{mm'}\,
\sum_l\vert C^{(1)}_l\vert^2 \equiv
\delta(k_y-k'_y)\delta(k_x-k'_x)\delta(k_--k'_-).
\label{orto}
\en

\subsection {The ``out" modes}

The metric (\ref{intmetric}) can be transformed, using the
Kruskal-Szekeres-like  coordinates defined by (\ref{defU'V'}) into  the
following form:
\be
ds^2=(1+\sin\xi)^2\;dU'\,dV' - {1\over 2ab}d\eta^2
-\cos^2\eta\;dx^2
\en
Near the horizon $\xi\to\pi/2$ the line element  is:
\be
ds^2=4\,dU'\,dV' - {1\over 2ab}d\eta^2 -\cos^2\eta\;dx^2
\en
One can see that at the horizon the line element possesses two null Killing
vector fields:  $\partial_{U'}$ and $\partial_{V'}$, so that  the ``out"-modes
will be taken as those with positive frequency with  respect to these Killing
vectors, and will have the form:
\be
\Phi (U',V',\eta,x) = e^{-i\omega_+U'-i\omega_-V'}\,\phi (\eta,x)
\en
The Klein-Gordon equation reduces to:
\be
\phi_{,\eta\eta}-\tan\eta\,\phi_{,\eta}+{\phi_{,xx}\over
2ab\cos^2\eta} +
{\omega_-\omega_+\over 2ab}\,\phi = 0
\en
Identifying the $x$ coordinate with the angular one, as in the subsection 3.3,
the general solution of this equation will be a lineal combination of the
modes:
\be
u^{out}_k(U',V',\eta,x)=
{1\over\sqrt{2\pi\,2\omega_\pm}}\,e^{-i\omega_+U'-i\omega_-
V'}\,
Y^m_l({\pi\over 2}-\eta,{x\over q}) \label{out}
\en
where:
\be
l(l+1) = {\omega_-\omega_+\over 2ab}.
\en
These modes are orthonormal at the horizon:
\be
(u^{out}_k,u^{out}_{k'}) = q^2 \delta_{ll'}\delta_{mm'}\delta
(\omega_\pm-\omega'_\pm).
\en
The $q^2$ factor insures the correct dimensions.

\section {Particle creation}

We now evaluate the Bogoliubov coeficients between the ``in"- modes and  the
``out"-modes at the horizon. The Bogoliubov transformation between these will
be:
\be
u^{in}_k = \sum_{k'} \left (\alpha_{kk'} u^{out}_{k'} +
                             \beta_{kk'} u^{out\ast}_{k'}\right )
\label{in-out}
\en
\be
u^{out}_k = \sum_{k'} \left (\alpha^{\ast}_{kk'} u^{in}_{k'} -
                             \beta_{kk'} u^{in\ast}_{k'}\right ),
\label{out-in}
\en
where $\alpha_{kk'}$ and $\beta_{kk'}$ can be found from:
\bear
\alpha_{kk'} &=& (u^{in}_k,u^{out}_{k'})\\
\beta_{kk'}  &=& -(u^{in}_k,u^{out\ast}_{k'})
\label{ab}
\enar
Substituting the expressions of  the modes (\ref{goodinmodes})  and (\ref{out})
to evaluate the scalar products at the horizon, we obtain:
\be
\alpha_{kk'}={q^2 C^{(1)}_{l'}\vert\hat k_y\vert
            \delta_{m,m'}
            \over 2\pi\sqrt{\vert k_y\vert\omega_{\pm}}}
            \Gamma [i\vert\hat k_y\vert ]
    \left\{\begin{array}{ll}
                 \left(i\hat\omega_+\right)^{-i\vert\hat
k_y\vert}&k_y\geq 0\\
                 \left(i\hat\omega_-\right)^{-i\vert\hat
k_y\vert}&k_y\leq 0
           \end{array}
    \right.\label{alpha}
\en
\be
\beta_{kk'}=-{q^2 (-1)^{\hat k'_x}\vert\hat k_y\vert C^{(1)}_{l'}
            \delta_{m,-m'}
            \over 2\pi\sqrt{\vert k_y\vert\omega_{\pm}}}
            \Gamma [i\vert\hat k_y\vert ]
    \left\{\begin{array}{ll}
           \left(-i\hat\omega_+\right)^{-i\vert\hat
k_y\vert}&k_y\geq 0\\
           \left(-i\hat\omega_-\right)^{-i\vert\hat
k_y\vert}&k_y\leq 0
           \end{array}
    \right.\label{beta}
\en

It can be easily shown that the relation between these coeficients  is:
\be
\vert\alpha_{kk'}\vert^2 = \exp{(2\pi\vert\hat k_y\vert)}\,
\vert\beta_{kk'}\vert^2, \label{thermal}
\en
and if one looks at the colliding wave problem in a time reversal  manner, i.e.
the waves running away from the initial caustic singularity  finally producing
a flat background region, then the exponential term of the equation
(\ref{thermal}) would give rise  to a thermal  spectrum of particles (the
number of particles as seen by the ``in"  observer when the field is in the
``out" vacuum).

In the colliding wave problem, one is interested to calculate the number of
``out" particles in the ``in" vacuum at the horizon. The  number of ``out"
particles with frequencies in a range between  $\omega_\pm$ and $\omega_\pm +
d\omega_\pm$, that the static ``out" observer  sees at the horizon if the field
is in the ``in" vacuum state $\vert 0,in>$ is  given by:
\be
N^{out}_{k'}\equiv a^{out\dagger}_{k'} a^{out}_{k'} = \int d^3 k
\vert\beta_{kk'}\vert^2 = {1\over q^2}\,\sum_{m}\sum_{l}\,\int
dk_y\, \vert\beta_{kk'}\vert^2.
\en
Introducing the Bogoliubov coeficients $\beta$ given by the equation
(\ref{beta}) one can get the following expression for the  number of created
particles:
\be
N^{out}_{\omega_{\pm}}= {q^3\over 2\pi}
                  {\delta_{m,-m'}\over 2\pi\hat\omega_{\pm}}
                \sum_l\,
                \int {dk \over e^k-1} \vert C^{(1)}_{l'}\vert^2
\label{number}
\en
where $k$ is a dimensionless variable defined by $k\equiv 2\pi  k_y$. The
coeficients $C^{(1)}_{l'}$ depend on $m,\,l$ and $k$ and can be in principle
evaluated explicitly by comparing the expression  given  by the equation
(\ref{general}) and the ``in" modes in the regions II (\ref{modesinII}) and III
at the wave fronts as well as imposing the orthonormalization relation
(\ref{orto}).

Comparing the expression (\ref{number}) with those obtained by  Yurtsever
\cite{yurtsever2} and Dorca and Verdaguer \cite{DV} in the cases  of pure
gravitational wave scattering, one can see that the results are  similar. The
equation (\ref{number}) is consistent with the long wavelength  limit of a
thermal distribution of scalar particles with a temperature given  by:
\be
T = {\hbar c\over k_{_B}}\,{1\over 2\pi q} = {\hbar c\over k_{_B}}\,
{\sqrt{2ab}\over 2\pi}.
\en
We have limited ourselves in this work to  the case of the neutral  particles.
In the case where the particles would have some charge, one  would expect the
spectrum of created particles to be characterised not only by a  temperature
but by a chemical potential as well. At any rate, one would not  expect results
qualitatively different from those obtained by Lapedes  \cite{lapedes} in the
Bertotti-Robinson universe.

An interesting question to address is whether the quantum field  theory has
some implications on the arrow of time in the plane wave  collisions, i.e. to
ask as to whether there is an entropy increase in one way or  another. Related
question, as well, would be as to whether there is a preferable  analytic
extension across the horizon: a possible extension could be a static  one or
the
time symmetric one as pointed in reference \cite{hayward}. One hopes that the
quantum field theory  could help to answer this question. These and some other
questions are  currently being considered by the authors and will be discussed
elsewhere.

\vskip 1cm
\noindent{\Large\bf Acknowledgements}\vskip 1cm

We are grateful to Enric Verdaguer for helpful suggestions and  discussions.
A.F. would like to thank Chaim Charach for many useful  conversations in the
past. M.A.P.S. is grateful to Miquel Dorca, Manuel A. Valle and  I\~nigo
Egusquiza for helpful discussions.

This work is supported by the Spanish Ministry of Education Grant  (CICYT) No.
PB-$93-0507$. M.A.P.S. is also supported by a Spanish Ministry of  Education
pre-doctoral fellowship No. AP92 $\;30619396$.


\end{document}